\documentclass[twocolumn,showpacs,preprintnumbers,amsmath,amssymb]{revtex4}

\usepackage{graphicx}
\usepackage{dcolumn}
\usepackage{bm}
\usepackage{braket}

\newcommand{\eye}{\mathbb I}

\begin{document}

\preprint{APS/123-QED}

\title{Preparation of pure and mixed polarization qubits and the direct measurement of figures of merit}
\author{R. B. A. Adamson}
\affiliation{%
Centre for Quantum Information \& Quantum Control and Institute for
Optical Sciences, Dept. of Physics, 60 St. George St., University of
Toronto, Toronto, ON, Canada, M5S 1A7
}%
\author{L. K. Shalm}
\affiliation{%
Centre for Quantum Information \& Quantum Control and Institute for
Optical Sciences, Dept. of Physics, 60 St. George St., University of
Toronto, Toronto, ON, Canada, M5S 1A7
}%

\author{A. M. Steinberg}
\affiliation{%
Centre for Quantum Information \& Quantum Control and Institute for
Optical Sciences, Dept. of Physics, 60 St. George St., University of
Toronto, Toronto, ON, Canada, M5S 1A7
}%

\date{\today}
\begin{abstract}
Non-classical joint measurements can hugely improve the efficiency
with which certain figures of merit of quantum systems are measured.
We use such a measurement to determine a particular figure of merit,
the purity, for a polarization qubit.  In the process we highlight
some of subtleties involved in common methods for generating
decoherence in quantum optics.
\end{abstract}

\maketitle

Quantum information science has the potential to dramatically
increase the speed of certain information processing tasks such as
factoring large numbers\cite{Shor1994}.  The superiority of quantum
information processors over classical ones arises in part because
the number of internal states of the processor increases
exponentially with the number of inputs and output bits, rather than
polynomially. This very property that makes quantum systems such
powerful information processors also makes them notoriously
difficult to characterize, since describing the state of a system at
a particular stage of a calculation requires exponentially more
measurements than there are input and output bits.  Even with the
small quantum information processors of ten or so qubits that have
been demonstrated so far\cite{Vand2001}, this characterization -
known technically as quantum state tomography\cite{Jame2001} - would
require up to a million independent measurements and rapidly becomes
impractical as systems become larger.

Given this problem, it is often desirable to describe the system in
terms of a few figures of merit that encapsulate the relevant
properties of the state for some particular application. The
simplest such figure of merit is the fidelity to the expected state
of the system\cite{MikeAndIke}. If the expected state is
$\ket{\phi}$, then one will measure the expectation value of the
projector $\mathbf{P}_\phi=\ket{\phi}\bra{\phi}$, and the fidelity
is given by $F=\text{Tr}\left\{\mathbf{P}_\phi \rho\right\}$, where
$\rho$ is the state of the system.  The fidelity measurement will
yield one if the system is in the expected state and less than one
if it is not. Other figures of merit measure more subtle properties
of the system. A partial list includes the purity\cite{MikeAndIke},
the Von Neumann entropy\cite{vonN1955}, the tangle\cite{Woot1998},
the concurrence\cite{Woot1998}, and the trace distance from another
state\cite{Gilc2005}. All of these figures of merit share the
property that they are non-linear functions of the density matrix as
opposed to fidelity which is linear in the density matrix. Whereas
the fidelity can be measured straight-forwardly as an expectation
value, these non-linear functions cannot be measured in this way.
Instead, the figure of merit is usually computed from the density
matrix. This presents an experimental problem because the number of
measurements required to determine the density matrix rises
exponentially with the size of the quantum system.

A means of circumventing this problem was proposed by Todd
Brun\cite{Brun2004} who showed how these non-linear functions can be
measured very efficiently with non-classical joint measurements, at
least in the special case that they are polynomial in the density
matrix. Brun demonstrated that an $m$th degree polynomial function
of the density matrix can be written as the expectation value of a
joint measurement performed on $m$ copies of the system described by
the same density matrix $\rho$. Since the density matrix describing
these $m$ copies can be written $\rho_\text{m}=\rho^{\otimes m}$,
expectations values that are linear in $\rho_\text{m}$ are $m$th
order polynomials in $\rho$. In particular, purity, defined as
\begin{equation}
\label{purity definition}
 P=\text{Tr}\left\{\rho^2\right\},
\end{equation}
being quadratic in $\rho$, can be measured directly as a joint
expectation value on two copies of a system.  Even non-polynomial
functions like the Von Neumann entropy can be measured by
approximating them with a truncated Taylor series in the density
matrix.

The purity is a useful figure of merit in many situations.  It is
directly related to the thermodynamic temperature of the ensemble
which can be easily calculated from it.  It can also be used as a
measure of entanglement of a particle with other systems, since for
a set of interacting systems each individual system will appear pure
when the overall state is separable, completely impure when the
state is maximally entangled and partially pure when the state is
partially entangled.  In this paper we will discuss applying Brun's
technique to the measurement of purity.

\begin{figure*}
  \centerline{
    \mbox{\includegraphics[width=\textwidth]{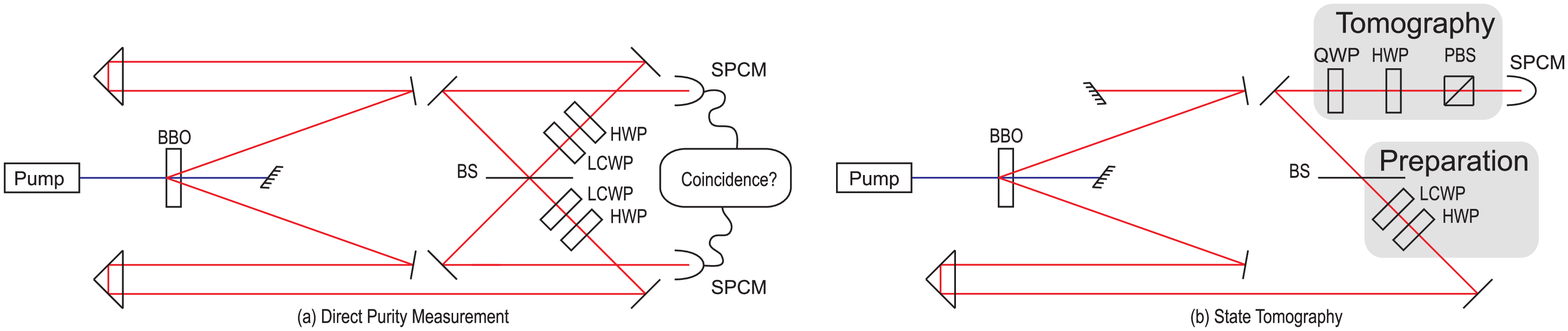}}
    }
  \caption{Experimental implementation of (a) the direct purity measurement and (b) quantum state
  tomography.
  Labels designate a 50/50 beamsplitter (BS), a non-linear $\beta$-Barium
  Borate (BBO) crystal, half waveplates (HWP), liquid-crystal variable waveplates (LCWP),
  single photon counting modules (SPCM) and a polarizing beamsplitter (PBS).
  A Type-I spontaneous parametric downconversion (SPDC) crystal produces pairs of H-polarized
  photons.  In (a), the same state $\rho$ is prepared in both arms
  and the beamsplitter acts as a singlet state filter.  In (b), a
  state is prepared in one of the arms and polarimetry is used to
  measure the density matrix of the state using a quarter
  waveplate, a half waveplate and a polarizing beamsplitter.}
  \label{apparatus}
  \end{figure*}

Joint measurements can result in an enormous reduction in the
resources required to measure non-linear figures of merit.   While
complete characterization scales exponentially with the size of the
system, the joint measurement used in the Brun technique is fixed by
the degree of the polynomial defining the figure of merit. A
10-qubit system requires over one million measurements to measure
the density matrix, but only a single joint measurement on pairs of
copies of the system to measure the purity.  The joint measurement
method for measuring purity was also applied experimentally in a
nuclear magnetic resonance system by Du et al\cite{Du2005}, and in
an entangled photon system by Bovino et al\cite{Bovi2005}.  Both
these systems were limited in the range of mixed states that they
were able to generate.  Recently, another figure of merit, the
concurrence, has also been directly measured in a photonic
system\cite{Walb2006}. Here we study the application of this
technique to a broad range of pure and mixed states and find that
the effectiveness of the approach depends crucially on the details
of how the state is prepared.


For an ensemble of single qubits described by a density matrix
\begin{equation}
\rho= \left(
\begin{matrix}
\rho_{00} & \rho_{01} \\
\rho_{10} & \rho_{11}\\
\end{matrix}
\right),
\end{equation}
the formula $P=\text{Tr}\left[\rho^2\right]$ can be expanded as a
second-order polynomial in the density matrix elements:
$P=\rho_{00}^2+\rho_{01}\rho_{10}+\rho_{10}\rho_{01}+\rho_{11}^2$.
According to Brun's result, the purity will be equal to the
expectation value of a two-particle operator $\bf A$ made by
replacing each quadratic term $\rho_{ij} \rho_{kl}$ in the sum  with
the two particle projector $\ket{i}\bra{j}\otimes\ket{k}\bra{l}$
such that
\begin{align}
\notag {\bf
A}=&\ket{0}\bra{0}\otimes\ket{0}\bra{0}+\ket{0}\bra{1}\otimes\ket{1}\bra{0}+\\
\notag
&\ket{1}\bra{0}\otimes\ket{0}\bra{1}+\ket{1}\bra{1}\otimes\ket{1}\bra{1}
\end{align}
Inserting the two-photon polarization identity operator $\eye_4$ and
grouping tensor products into two photon states, we obtain
\begin{align}
{\bf
A}=&\eye_4-\ket{01}\bra{01}-\ket{10}\bra{10}+\ket{01}\bra{10}+\ket{10}\bra{01}\\
=&\eye_4-2\ket{\psi^-}\bra{\psi^-},
\end{align}
where $\ket{\psi^\pm} \equiv\left(
\ket{01}\pm\ket{10}\right)/\sqrt{2}$. Applying $P=\left<{\bf
A}\right>=\text{Tr}\left\{{\bf A}\rho\otimes\rho\right\}$, we obtain
\begin{align}
\notag
P&=\text{Tr}\left\{\left(\eye_4-2\ket{\psi^-}\bra{\psi^-}\right)\rho\otimes\rho\right\}\\
\label{purityformula} &=1-2\bra{\psi^-}\rho\otimes\rho\ket{\psi^-}
\end{align}
Thus the purity can be obtained by a single measurement on two
particles, namely a projection onto the singlet state
$\ket{\psi^-}$. This fact can be intuitively understood by realizing
that a projection onto $\ket{\psi^-}$ implements a measurement of
permutation symmetry, since the two qubit space divides into a
symmetric subspace spanned by
$\left\{\ket{00},\ket{11},\ket{\psi^+}\right\}$ and the orthogonal,
antisymmetric state $\ket{\psi^-}$.  If a state $\rho$ is pure, then
the state $\rho \otimes \rho$ is manifestly permutation invariant,
and therefore has no antisymmetric component.  In the other limit of
a completely mixed state $\rho=\frac{1}{2} \eye_2$,
$\rho\otimes\rho=\frac{1}{4}\eye_4$ with a projection onto the
singlet state (and any other state) of 0.25 and hence a purity of
0.5.  For states of intermediate purity $\rho \otimes \rho$ can be
decomposed into antisymmetric and symmetric parts, the relative
weights of which determine the purity.

In quantum optics, $\ket{\psi^-}\bra{\psi^-}$ can be measured using
the Hong-Ou-Mandel (HOM) effect \cite{HOM1987}. The effect occurs
when two photons are made indistinguishable in all physical
properties except polarization and then sent into the two input
ports of a beamsplitter so as to arrive at the same time.  Photons
in a permutation-symmetric polarization state such as $\ket{HH}$ or
$\ket{\psi^+}$ will always leave the beamsplitter in the same port,
whereas if the incoming photons are in the permutation-antisymmetric
state $\ket{\psi^-}$ then they will always leave the beamsplitter in
opposite ports.  If we measure the rate of coincident firings of
detectors at the two output ports we will have filtered for the
singlet state $\ket{\psi^-}$, thereby, through
eq.($\ref{purityformula}$), measuring the purity. This technique of
using the HOM effect as a singlet state filter has been employed in
many important quantum optics experiments including the
demonstration of teleportation\cite{Bouw1997}.  Its applicability
for this task is discussed by Mitchell et al\cite{Mitc2003} who
fully characterized the filtering process using quantum process
tomography and by Kim and Grice\cite{Kim2003} who note some of its
limitations.

\begin{figure}[!t]
  \centerline{
    \mbox{\includegraphics[width=\columnwidth]{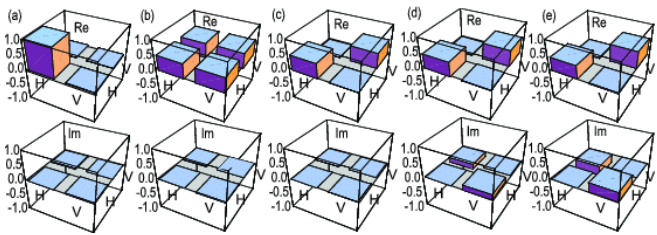}}
  }
  \caption{Experimentally obtained density matrices for various single-photon polarization
  states.
  (a) pure horizontal $\ket{H}$
  (b) pure diagonal $\ket{+}$
  (c) equal mixture of $\ket{+}$ and $\ket{-}$
  (d) equal mixture of $\ket{+}$, $\ket{-}$ and $\ket{R}=(\ket{H}-i
  \ket{V})/\sqrt{2}$
  (e) Of mixture of states of the form $(\ket{H}+e^{i \phi}\ket{V})/\sqrt{2}$ with $\phi$ distributed
  equally over $\left[0, \pi\right]$}
  \label{uncorrelated data}
  \end{figure}

The experimental implementation of the purity measurement was
carried out using the apparatus shown in Fig.\ref{apparatus}(a).  A
BBO crystal cut for Type-I phase-matching was pumped with a 28-mW,
405-nm diode laser, producing pairs of horizontally polarized
photons over a wide bandwidth.  The photons were sent through 10-nm
interference filters to reduce spectral differences between them.  A
corner prism mounted on a motorized translation stage allowed
control of the relative delay in the two arms. The photons passed
through a half waveplate and a liquid crystal variable waveplate
with its axis aligned with the horizontal.  The photons then arrived
at the beamsplitter which implemented the joint measurement before
passing to the detector.  Ideally the beamsplitter should act as a
perfect singlet state filter and never allow the photons to leave in
opposite ports when the input polarization states are the same.  In
reality, the visibility of the two-photon interference was limited
to $0.90\pm0.03$ by the unequal reflection and transmission
coefficients of the beamsplitter and by small alignment errors.  As
a result, the measurement actually implemented was
${\bf{P}}_\text{actual}=0.10 I+0.90 \ket{\psi^-}\bra{\psi^-}$ rather
than ${\bf{P}}_\text{ideal}=\ket{\psi^-}\bra{\psi^-}$. The tabulated
purities were calculated by taking account of this modified
measurement:
$\left<{\bf{P}}_\text{ideal}\right>=\frac{\left<{\bf{P}}_\text{actual}\right>-0.1}{0.9}$

This is not quite equivalent to the measurement that might be made
with an ideal singlet state filter. Since in practice
$\left<{\bf{P}}_\text{actual}\right>$ represents a measured
expectation value obtained from a finite number of measurements it
will have some shot noise associated with it. Consequently the
estimate of $\left<{\bf{P}}_\text{ideal}\right>$ will also have some
shot noise and thus for a pure state there is a finite probability
of obtaining a value for the singlet state projection of less than
zero and hence a purity greater than one. Similarly for mixed states
there is a finite probability of obtaining a purity less than $1/2$
with a finite number of measurements.  The probability of obtaining
these unphysical results goes to zero as $1/\sqrt{N}$ in the limit
of a large number of measurements $N$, and it is only in this limit
that our expression for the purity may be considered exact. This is
in keeping with the well-known fact in statistics that an exact
estimate of an expectation value requires an infinite number of
measurements.

Impure states were created by applying random polarization phase
shifts with the liquid crystal waveplates. Thus polarization was
correlated to a pseudo-random number generator rather than to some
traced-over degree of freedom. As far as polarization measurements
are concerned, there is no observable difference between impure
states generated with this technique and those generated by some
more complicated method such as loss of a single photon from an
entangled pair\cite{Pete2004}, although, as will be discussed later,
there is a difference between this approach and the generation of
decoherence by coupling to another degree of freedom of the photon.

For comparison, the purity was also determined by measuring the
density matrix using quantum state tomography and applying the
formula $P=\text{Tr}\{\rho^2\}$.  Quantum state tomography was
performed by blocking one of the photons and performing projective
measurements on the other photon. Background counts due to detector
dark counts and residual light were subtracted from the data before
reconstructing the density matrix.

\begin{table}[!t]
\begin{tabular}
[c]{c|ccc}%
&&{\bf Purities}&\\
{\bf State} &{\bf Direct}&{\bf Tomographic}&{\bf Theoretical}\\
\hline\hline\\
$\ket{H}$&$1.00\pm0.03$&$1.00\pm0.01$&$1$\\
\hline\\
$\ket{+}$&$0.99\pm0.03$&$0.98\pm0.01$&$1$\\
\hline
Equal mixture\\
$\ket{+}$, $\ket{-}$&$0.52\pm0.01$&$0.50\pm0.01$&$0.5$\\
\hline\\
Equal mixture\\
$\ket{+}$, $\ket{-}$, $\ket{R}$&$0.568\pm0.008$&$0.56\pm0.01$&$5/9\approx0.5556$\\
\hline\\
$\ket{H}+e^{i\phi}\ket{V}$,\\
 $\phi \in \left[0,\pi\right]$ &$0.72\pm0.01$&$0.70\pm0.01$&$0.5+\frac{2}{\pi^2}\approx0.7026$\\
 \hline\\

\end{tabular}
\centering \caption{The purities measured for five states using the
direct joint purity measurement and a full characterization followed
by a calculation of $\text{Tr}\{\rho^2\}$.  The stated errors arise
from counting statistics and the statistics associated with the
random state selection.} \label{results}
\end{table}

Table \ref{results} shows the results of state tomography and direct
purity measurement for a variety of states of varying purity. A
number of non-decohering and decohering preparation processes were
performed.  If no preparation was done the state was left in
$\ket{H}$ as in Fig.\ref{uncorrelated data}(a) with essentially unit
purity measured with both methods.  Fig. \ref{uncorrelated data}(b)
shows the state after a unitary rotation to the state $\ket{+}$
where we have used the notation
$\ket{\pm}\equiv\ket{H}\pm\ket{V}/\sqrt{2}$. This rotation has no
effect on the purity as measured with either method since purity is
an invariant under unitary operations. Fig. \ref{uncorrelated
data}(c) shows the completely mixed state $\rho=\frac{1}{2}\eye_2$
obtained by randomly applying a phase shift of $0$ or $\pi$ with the
liquid crystal waveplates to the state $\ket{+}$.  The purity
measured with either method is consistent with the theoretical value
of $0.5$ or completely mixed. Fig. \ref{uncorrelated data}(d) shows
a state made by randomly applying either a $0$, $\pi$ or $\pi/2$
phase shift to the state $\ket{+}$. The purity is consistent with
the theoretical value of $5/9$. Finally, Fig. \ref{uncorrelated
data}(e) shows a state created by selecting randomly from the
continuum of phases in the range $\left[0,\pi\right]$. Again, the
measured purity was consistent with the theoretical value of
$0.5+\frac{2}{\pi^2}$.

While the joint measurement technique correctly measures the purity
in these cases, caution must be exercised in using it since the
method depends crucially on the assumption that the couplings to the
environment that cause the reductions in purity in the two copies of
the state be uncorrelated. In the case where the couplings to the
environment are perfectly correlated, any given two copies of the
state will be in the same pure state at any given moment, and the
direct purity measurement will indicate unit purity.  In this
experiment, correlated decoherence was achieved by applying phase
shifts to the two photons with liquid crystal waveplate so as to
always give the two photons the same birefringent phase, even as the
value of this phase varied randomly over time. In this situation the
density matrix for either individual photon is mixed since the phase
shifts are random, but at any given moment the two photons are in
the same state, and hence their singlet state projection is always
zero.

The purity was measured directly for the three mixed states (Figs.
2c, 2d and 2e) generated with the mixing completely correlated for
the two copies of the system. All measured purities were consistent
with unit purity.  This demonstrates that the joint measurement
technique relies on the assumption that the decoherence processes
are independent for all the particles being jointly measured and
fails to work when this assumption doesn't hold.

The quantum optics implementation of this technique depends in
another way on the details of how mixed photon states are prepared.
 A common technique for creating impure photon
polarization states is to create a correlation between polarization
and some other photon degree of freedom such as frequency or spatial
mode\cite{Pete2004} which is subsequently ignored or traced over in
the measurement. However, for the HOM effect to perform as a
polarization singlet state filter, the photons must be
indistinguishable in all degrees of freedom other than polarization,
and this will not be the case if extra degrees of freedom become
correlated to polarization.
\begin{figure}[!t]
  \centerline{
    \mbox{\includegraphics[width=\columnwidth]{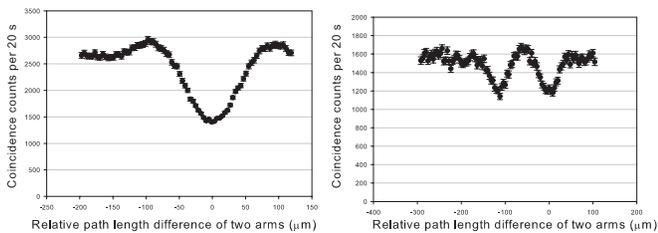}}
  }
  \caption{(a) The HOM dip expected for a maximally mixed qubit state
  (b) Observed data when such a state is created by introducing a large birefringent group delay between horizontal and veritcal polarizations. }
  \label{HOM data}
  \end{figure}

To demonstrate this fact the same experiment was done using the
method of \cite{Pete2004} to create impurity.  In this method,
impurity is obtained by creating a frequency-dependent birefringent
phase shift (or equivalently, introducing a constant group delay
between the horizontal and vertical components).  Since the photon
measurement is insensitive to frequency, this correlation is traced
over, resulting in a loss of coherence between the horizontal and
vertical components of the density matrix.  In order to create the
maximally mixed state $\rho=\frac{1}{2}\eye_2$ a 20mm piece of
quartz was introduced into each beam with the fast and slow axes of
the two crystals relatively rotated by $90^\text{o}$ so that the
vertical polarization component was advanced in one arm relative to
the horizontal component and delayed in the other. Diagonally
polarized light was then sent into each arm. State tomography on the
individual photons resulted in the expected density matrix
$\rho=\frac{1}{2}\eye_2$, but the HOM visibility did not result in
the correct value for the purity.  As the path length difference in
the two arms was scanned, rather than observing a single HOM dip
yielding a purity of 0.5 as in Fig.\ref{HOM data}(a), two dips were
observed as in Fig.\ref{HOM data}(b), either of which would have
implied a purity of 0.25. The discrepancy can be accounted for by
noting that after the delay is introduced, a complete description of
the state of each photon involves not only polarization but also the
relative delay between the horizontal and vertical components. The
state space becomes four-dimensional, being spanned by
$\left\{\ket{H,\text{early}},\ket{H,\text{late}},\ket{V,\text{early}},\ket{V,\text{late}}\right\}$.
Coincidences are suppressed when either both photons arrive late
with the same polarization or both arrive early with the same
polarization. When either the polarizations or the arrival times are
different the photons cause a coincidence 50\% of the time.  Since
out of 16 possible combinations of arrival time and polarization
only four are removed by singlet state filtering the dip visibility
is 25\% as observed. This result demonstrates the danger of thinking
of the HOM effect as acting as a polarization singlet state filter.
Rather the HOM effect is a symmetry filter over all the properties
of the photon, even those that one would prefer that it ignore.

For completeness we also looked at the case where the crystals were
oriented so as to delay the same polarization in the two arms. In
that case the interaction with the `environment' is correlated and
the photons in the two arms are always in the same state in the
four-dimensional state space. Just as when the correlated mixed
states were created with liquid crystals waveplates, the joint
measurement of the purity for correlated delays gives a purity of
one even though the polarization density matrix measured with
tomography is mixed.

We have demonstrated the direct measurement of the purity of a
single qubit through a joint measurement on two particles drawn from
an ensemble.  Such measurements are much more efficient at
determining non-linear figures of merit such as purity than the
common method of measuring the density matrix and calculating the
figure of merit from it, particularly in large Hilbert spaces.
Furthermore, we have shown that the method depends critically on the
properties of the decoherence process generating a mixture of
states. Joint measurements on a mixed state where the decoherence is
correlated between the measured photons generate the same
experimental signature as no decoherence at all, while a widely-used
technique for preparing decohered states fails because of the
behaviour of the HOM effect when used as a singlet state filter.
These caveats aside, joint measurements still provide an efficient
means of characterizing quantum states that may find application in
thermometry, quantum error correction and the characterization of
quantum devices.

The authors would like to thank Jeff Lundeen for useful discussions,
and Jan Henneberger and Alan Stummer for technical assistance.  This
work was supported by the Natural Science and Engineering Research
Council of Canada, the DARPA QuIST program managed by the US AFOSR
(F49620-01-1-0468), Photonics Research Ontario, the Canadian
Institute for Photonics Innovation and the Canadian Institute for
Advanced Research. RBAA is funded by the Walter C. Sumner
Foundation.

\bibliography{paper}

\begin{thebibliography}{16}
\expandafter\ifx\csname natexlab\endcsname\relax\def\natexlab#1{#1}\fi
\expandafter\ifx\csname bibnamefont\endcsname\relax
  \def\bibnamefont#1{#1}\fi
\expandafter\ifx\csname bibfnamefont\endcsname\relax
  \def\bibfnamefont#1{#1}\fi
\expandafter\ifx\csname citenamefont\endcsname\relax
  \def\citenamefont#1{#1}\fi
\expandafter\ifx\csname url\endcsname\relax
  \def\url#1{\texttt{#1}}\fi
\expandafter\ifx\csname urlprefix\endcsname\relax\def\urlprefix{URL }\fi
\providecommand{\bibinfo}[2]{#2}
\providecommand{\eprint}[2][]{\url{#2}}

\bibitem[{\citenamefont{Shor}(1994)}]{Shor1994}
\bibinfo{author}{\bibfnamefont{P.~W.} \bibnamefont{Shor}},
  \bibinfo{journal}{Foundations of Computer Science, 1994 Proceedings., 35th
  Annual Symposium on} pp. \bibinfo{pages}{124--134} (\bibinfo{year}{1994}).

\bibitem[{\citenamefont{Vandersypen et~al.}(2001)\citenamefont{Vandersypen,
  Steffen, Breyta, Yannoni, and Chuang}}]{Vand2001}
\bibinfo{author}{\bibfnamefont{L.~M.~K.} \bibnamefont{Vandersypen}},
  \bibinfo{author}{\bibfnamefont{M.}~\bibnamefont{Steffen}},
  \bibinfo{author}{\bibfnamefont{G.}~\bibnamefont{Breyta}},
  \bibinfo{author}{\bibfnamefont{C.~S.} \bibnamefont{Yannoni}},
  \bibnamefont{and} \bibinfo{author}{\bibfnamefont{M.~H. S. . I.~L.}
  \bibnamefont{Chuang}}, \bibinfo{journal}{Nature}
  \textbf{\bibinfo{volume}{414}}, \bibinfo{pages}{883} (\bibinfo{year}{2001}).

\bibitem[{\citenamefont{James et~al.}(2001)\citenamefont{James, Kwiat, Munro,
  and White}}]{Jame2001}
\bibinfo{author}{\bibfnamefont{D.~F.~V.} \bibnamefont{James}},
  \bibinfo{author}{\bibfnamefont{P.~G.} \bibnamefont{Kwiat}},
  \bibinfo{author}{\bibfnamefont{W.~J.} \bibnamefont{Munro}}, \bibnamefont{and}
  \bibinfo{author}{\bibfnamefont{A.~G.} \bibnamefont{White}},
  \bibinfo{journal}{Phys. Rev. A.} \textbf{\bibinfo{volume}{64}},
  \bibinfo{pages}{052312} (\bibinfo{year}{2001}).

\bibitem[{\citenamefont{Nielsen and Chuang}(2000)}]{MikeAndIke}
\bibinfo{author}{\bibfnamefont{M.~A.} \bibnamefont{Nielsen}} \bibnamefont{and}
  \bibinfo{author}{\bibfnamefont{I.~L.} \bibnamefont{Chuang}},
  \emph{\bibinfo{title}{Quantum Computation and Quantum Information}}
  (\bibinfo{publisher}{Cambridge University Press},
  \bibinfo{address}{Cambridge, UK}, \bibinfo{year}{2000}).

\bibitem[{\citenamefont{von Neumann}(1955)}]{vonN1955}
\bibinfo{author}{\bibfnamefont{J.}~\bibnamefont{von Neumann}},
  \emph{\bibinfo{title}{Mathematical Foundations of Quantum Mechanics}}
  (\bibinfo{publisher}{Princeton University Press},
  \bibinfo{address}{Princeton, NJ}, \bibinfo{year}{1955}).

\bibitem[{\citenamefont{Wootters}(1998)}]{Woot1998}
\bibinfo{author}{\bibfnamefont{W.~K.} \bibnamefont{Wootters}},
  \bibinfo{journal}{Phys. Rev. Lett.} \textbf{\bibinfo{volume}{80}},
  \bibinfo{pages}{2245} (\bibinfo{year}{1998}).

\bibitem[{\citenamefont{Gilchrist et~al.}(2005)\citenamefont{Gilchrist,
  Langford, and Nielsen}}]{Gilc2005}
\bibinfo{author}{\bibfnamefont{A.}~\bibnamefont{Gilchrist}},
  \bibinfo{author}{\bibfnamefont{N.~K.} \bibnamefont{Langford}},
  \bibnamefont{and} \bibinfo{author}{\bibfnamefont{M.~A.}
  \bibnamefont{Nielsen}}, \bibinfo{journal}{Phys. Rev. A}
  \textbf{\bibinfo{volume}{71}}, \bibinfo{pages}{062310}
  (\bibinfo{year}{2005}).

\bibitem[{\citenamefont{Brun}(2004)}]{Brun2004}
\bibinfo{author}{\bibfnamefont{T.~A.} \bibnamefont{Brun}},
  \bibinfo{journal}{Quantum Information and Computation}
  \textbf{\bibinfo{volume}{4}}, \bibinfo{pages}{401} (\bibinfo{year}{2004}),
  \eprint{quant-ph/0401067}.

\bibitem[{\citenamefont{Du et~al.}(2005)\citenamefont{Du, Zou, Oi, Peng, Kwek,
  Oh, and Ekert}}]{Du2005}
\bibinfo{author}{\bibfnamefont{J.}~\bibnamefont{Du}},
  \bibinfo{author}{\bibfnamefont{P.}~\bibnamefont{Zou}},
  \bibinfo{author}{\bibfnamefont{D.~K.~L.} \bibnamefont{Oi}},
  \bibinfo{author}{\bibfnamefont{X.}~\bibnamefont{Peng}},
  \bibinfo{author}{\bibfnamefont{L.~C.} \bibnamefont{Kwek}},
  \bibinfo{author}{\bibfnamefont{C.~H.} \bibnamefont{Oh}}, \bibnamefont{and}
  \bibinfo{author}{\bibfnamefont{A.}~\bibnamefont{Ekert}},
  \emph{\bibinfo{title}{Experimental demonstration of quantum state multi-meter
  and one-qubit fingerprinting in a single quantum device}}
  (\bibinfo{year}{2005}), \eprint{quant-ph/0411180}.

\bibitem[{\citenamefont{Bovino et~al.}(2005)\citenamefont{Bovino, Castagnoli,
  Ekert, Horodecki, Alves, and Sergienko}}]{Bovi2005}
\bibinfo{author}{\bibfnamefont{F.~A.} \bibnamefont{Bovino}},
  \bibinfo{author}{\bibfnamefont{G.}~\bibnamefont{Castagnoli}},
  \bibinfo{author}{\bibfnamefont{A.}~\bibnamefont{Ekert}},
  \bibinfo{author}{\bibfnamefont{P.}~\bibnamefont{Horodecki}},
  \bibinfo{author}{\bibfnamefont{C.~M.} \bibnamefont{Alves}}, \bibnamefont{and}
  \bibinfo{author}{\bibfnamefont{A.~V.} \bibnamefont{Sergienko}},
  \bibinfo{journal}{Phys. Rev. Lett.} \textbf{\bibinfo{volume}{95}},
  \bibinfo{pages}{240407} (\bibinfo{year}{2005}).

\bibitem[{\citenamefont{Walborn et~al.}(2006)\citenamefont{Walborn, Ribeiro,
  Davidovich, Mintert, and Buchleitner}}]{Walb2006}
\bibinfo{author}{\bibfnamefont{S.~P.} \bibnamefont{Walborn}},
  \bibinfo{author}{\bibfnamefont{P.~H.~S.} \bibnamefont{Ribeiro}},
  \bibinfo{author}{\bibfnamefont{L.}~\bibnamefont{Davidovich}},
  \bibinfo{author}{\bibfnamefont{F.}~\bibnamefont{Mintert}}, \bibnamefont{and}
  \bibinfo{author}{\bibfnamefont{A.}~\bibnamefont{Buchleitner}},
  \bibinfo{journal}{Nature} \textbf{\bibinfo{volume}{440}},
  \bibinfo{pages}{1022} (\bibinfo{year}{2006}).

\bibitem[{\citenamefont{Hong et~al.}(1987)\citenamefont{Hong, Ou, and
  Mandel}}]{HOM1987}
\bibinfo{author}{\bibfnamefont{C.~K.} \bibnamefont{Hong}},
  \bibinfo{author}{\bibfnamefont{Z.~Y.} \bibnamefont{Ou}}, \bibnamefont{and}
  \bibinfo{author}{\bibfnamefont{L.}~\bibnamefont{Mandel}},
  \bibinfo{journal}{Phys. Rev. Lett.} \textbf{\bibinfo{volume}{59}},
  \bibinfo{pages}{2044} (\bibinfo{year}{1987}).

\bibitem[{\citenamefont{Bouwmeester et~al.}(1997)\citenamefont{Bouwmeester,
  Pan, Mattle, Eibl, Weinfurter, and Zeilinger}}]{Bouw1997}
\bibinfo{author}{\bibfnamefont{D.}~\bibnamefont{Bouwmeester}},
  \bibinfo{author}{\bibfnamefont{J.-W.} \bibnamefont{Pan}},
  \bibinfo{author}{\bibfnamefont{K.}~\bibnamefont{Mattle}},
  \bibinfo{author}{\bibfnamefont{M.}~\bibnamefont{Eibl}},
  \bibinfo{author}{\bibfnamefont{H.}~\bibnamefont{Weinfurter}},
  \bibnamefont{and}
  \bibinfo{author}{\bibfnamefont{A.}~\bibnamefont{Zeilinger}},
  \bibinfo{journal}{Nature} \textbf{\bibinfo{volume}{390}},
  \bibinfo{pages}{575} (\bibinfo{year}{1997}).

\bibitem[{\citenamefont{Mitchell et~al.}(2003)\citenamefont{Mitchell, Ellenor,
  Schneider, and Steinberg}}]{Mitc2003}
\bibinfo{author}{\bibfnamefont{M.~W.} \bibnamefont{Mitchell}},
  \bibinfo{author}{\bibfnamefont{C.~W.} \bibnamefont{Ellenor}},
  \bibinfo{author}{\bibfnamefont{S.}~\bibnamefont{Schneider}},
  \bibnamefont{and} \bibinfo{author}{\bibfnamefont{A.~M.}
  \bibnamefont{Steinberg}}, \bibinfo{journal}{Phys. Rev. Lett.}
  \textbf{\bibinfo{volume}{91}}, \bibinfo{pages}{120402}
  (\bibinfo{year}{2003}).

\bibitem[{\citenamefont{Kim and Grice}(2003)}]{Kim2003}
\bibinfo{author}{\bibfnamefont{Y.-H.} \bibnamefont{Kim}} \bibnamefont{and}
  \bibinfo{author}{\bibfnamefont{W.~P.} \bibnamefont{Grice}},
  \bibinfo{journal}{Phys. Rev. A} \textbf{\bibinfo{volume}{68}},
  \bibinfo{pages}{062305} (\bibinfo{year}{2003}).

\bibitem[{\citenamefont{Peters et~al.}(2004)\citenamefont{Peters, Altepeter,
  Branning, Jeffrey, Wei, and Kwiat}}]{Pete2004}
\bibinfo{author}{\bibfnamefont{N.~A.} \bibnamefont{Peters}},
  \bibinfo{author}{\bibfnamefont{J.~B.} \bibnamefont{Altepeter}},
  \bibinfo{author}{\bibfnamefont{D.~A.} \bibnamefont{Branning}},
  \bibinfo{author}{\bibfnamefont{E.~R.} \bibnamefont{Jeffrey}},
  \bibinfo{author}{\bibfnamefont{T.-C.} \bibnamefont{Wei}}, \bibnamefont{and}
  \bibinfo{author}{\bibfnamefont{P.~G.} \bibnamefont{Kwiat}},
  \bibinfo{journal}{Phys. Rev. Lett.} \textbf{\bibinfo{volume}{92}},
  \bibinfo{pages}{133601} (\bibinfo{year}{2004}).

\end{thebibliography}

\end{document}